\documentstyle[12pt]{article}
\begin{document}
\title{Mona Lisa - ineffable smile of quantum mechanics}
\author{Slobodan Prvanovi\'c
\\
Institute of Physics, P.O. Box 57, 11080 Belgrade, Serbia}
\date{}
\maketitle
\begin{abstract}
The portrait of Mona Lisa is scrutinized with reference to quantum
mechanics. The elements of different expressions are firstly
recognized on her face. The contradictory details are then
classified in two pictures that, undoubtedly representing distinct
moods, confirm dichotomous character of the original. Consecutive
discussion has lead to conclusion that the mysterious state Mona
Lisa is in actually is coherent mixture - superposition, of
cheerfulness and sadness.
\end{abstract}

State of the physical system is among the most important concepts
of quantum mechanics. Being the primitive concept of the theory,
it is usually left undefined. However, by state we mean a list of
some relevant characteristics of the system in question. More or
less exact information about the quantities pertaining to the
system are on this list, telling us how it is prepared. For
operational reasons, state of the quantum mechanical system is
represented by some vector of the Hilbert space. Precisely, rays
represent states, while quantities are represented by the
Hermitian operators acting in this Hilbert space of states.

The spectral decomposition of the Hermitian operator, in the simplest
possible case when the spectrum is discrete and nondegenerate, reads:
\begin{equation}
\hat A = \sum _{i=1}^N a_i \vert a_i \rangle \langle a_i \vert ,
\end{equation}
where $N$ is dimension of Hilbert space. In this expression we have
used Dirac notation: $\vert a_i \rangle \langle a_i \vert$ is the
projector on the vector $\vert a_i \rangle$ - a formal representative
of the state. The real numbers $a_i$, being the eigenvalues -
characteristic values, of the operator $\hat A$, represent possible
numerical realizations of some quantity (which is represented above
by $\hat A$). Here, we are interested only in one quantity -
observable, and to each of its quantitative realizations $a_i$ there
corresponds appropriate state of the system described by $\vert a_i
\rangle$ and projector $\vert a_i \rangle \langle a_i \vert$.

Measurement is a process of acquisition of knowledge about the value of
some observable and about the state of the system. Measurements are
performed by some specially designed and, usually complicated physical
systems - apparatus. They interact with measured systems and, on the
other side, display numbers as results. And as the result of particular
measurement of $\hat A$, one  can get only some eigenvalue $a_i$.
Apparatus, or part of it, has to be classical mechanical system.
Otherwise, observer would not be able to obtain understandable
information about the value of measured quantity, which would be
against the purpose of apparatus.

In classical theory, when one performs the same experiment on equally
prepared systems whose state is completely known, one always obtains
the same results, {\it i.e.}, there is no probability argument. Quantum
mechanics, on the other hand, is essentially probabilistic theory.
Probability is intrinsic characteristic of this theory since there are
situations when one performs the same experiment on equally prepared
systems in completely known state and obtains different outcomes. In
such cases, it is impossible to predict which $a_i$ will result from
particular measurement. Knowledge of $\vert \psi \rangle$ enables
prediction of probability of finding system in the state $\vert a_i
\rangle$ only. It is given by:
\begin{equation}
\langle \psi \vert a_i \rangle \langle a_i \vert \psi \rangle = {\rm
Tr} (\vert \psi \rangle \langle \psi \vert \cdot \vert a_i \rangle
\langle a_i \vert ) ,
\end{equation}
where, according to the Dirac notation, $\langle \ \ \vert \ \ \rangle$
stands for the scalar product of involved vectors. This {\it a priori}
calculated probability matches with the relative frequency of $a_i$
(number of occurrences of this event divided by the total number of
performed measurements) if the number of performed experiments tends to
infinity. The set of equal, but independent systems used in these
experiments are called ensemble. More on the formalism of quantum
mechanics one can find in [1-2].

In case when the system is in state $\vert \psi \rangle \ne \vert a_i
\rangle$, while $\hat A$ is being measured, more than one outcome will
happen due to nonvanishing probability. On the other hand, these
probabilities are unavoidable, they do not follow from some observers
fault or imperfection of instruments. Their existence is not due to the
subjective, but to objective reasons.

Of course, as in classical mechanics, observer can be somewhat ignorant
regarding the state of quantum system. In this way additional
probability argument is being introduced into play. But, this incomplete
information probability is avoidable in principle, so it has to be
distinguished from the above-discussed one.

Formalism of quantum mechanics enables accurate reflection of all
possibilities when the preparation of a system is under question.
Firstly, concepts of pure and mixed states are distinguished. Pure
state is the one that is completely known and for which ensemble
of systems cannot be divided into subsets of differently prepared
systems. Otherwise, if we are not in position to  control
precisely preparation of systems or ensemble splits into
inequivalent subensembles, we are talking about mixed state.
Secondly, in order to enlighten difference between these types,
instead of vectors, operators are used to represent states. Then,
pure state is given by projector, {\it e. g.}, $\vert \psi \rangle
\langle \psi \vert$. It is obviously in one-to-one correspondence
with $\vert \psi \rangle$. In case of mixed or impure state we
know probability distribution over pure states: we know that there
is probability $w_1$ that system is in $\vert \psi _1 \rangle
\langle \psi _1 \vert$, $w_2$ that it is in $\vert \psi _2 \rangle
\langle \psi _2 \vert$ and so on. Standard notation for mixed
state is $\hat \rho = \sum _{j=1}^n w_j \vert \psi _j \rangle
\langle \psi _j \vert$, ($n\le N$). Of course, for probabilities
(weights) $w_j$ it holds $\sum _{j=1}^n w_j =1$. If the operator
is idempotent($\hat \rho ^2 = \hat \rho$) then the corresponding
state is pure while, otherwise, it is mixed.

For our purpose, it is interesting to distinguish coherent from
noncoherent mixtures. Despite of its name, coherent mixture is pure
state, while noncoherent mixture is proper mixed state. Difference is
best seen on concrete example. Coherent mixture of, say $\vert a_1
\rangle$ and $\vert a_2 \rangle$ is $\vert \psi \rangle = \sum _{i=1}^2
c_i \vert a_i \rangle$ or, in the operator form, it is given by:
\begin{equation}
\sum _{i=1}^2 \sum _{j=1}^2 c_i c_j ^* \vert a_i \rangle \langle a_j
\vert .
\end{equation}
Noncoherent mixture of the same states is:
\begin{equation}
\sum _{i=1}^2 w_i  \vert a_i \rangle \langle a_i \vert .
\end{equation}
Former state is pure and latter is not. Purity of coherent mixture
rests on presence of the so-called off-diagonal, $i\ne j$, elements.
Due to these terms, operator (3) is projector (idempotent).  All systems
in the ensemble represented by (3) are in the same state - $\vert \psi
\rangle$. For (4) situation is different; fraction of the ensemble,
equal to $w_1$, is in $\vert a_i \rangle$ and the rest is in $\vert a_2
\rangle$. Coefficients $c_i$, appearing in (3), are probability
amplitudes, while $w_i$ of (4) are probabilities of appropriate states.
Probability to obtain $a_i$, when system is in state (3), is $c_i
c_i^*$, while for (4) this probability is $w_i$ (these two
probabilities might be equal for particular choice of $c_i$ and $w_i$).
Occurrence of $\vert a_i \rangle$ in case of coherent mixture is
confirmation of intrinsic probability and since this state is assumed
to be linear combination of the eigenstates of measured observable, {\it
i.e.}, $\vert \psi \rangle = \sum _{i=1}^N c_i \vert a_i \rangle$, one
finds that intrinsic probability of quantum mechanics is closely
related to the superposition principle.

It could be said that, according to the superposition principle, if
$\vert a_i \rangle$ are allowed states of quantum system, then such is
every linear combination $\vert \psi \rangle = \sum _{i=1}^N c_i \vert
a_i \rangle$ where, of course, choice of $c_i$ ensures that $\vert \psi
\rangle$ has the norm equal to one ($\langle \psi \vert \psi \rangle
=1$). When system is in state $\sum _{i=1}^N c_i \vert a_i \rangle$, it
is not in one of $\vert a_i \rangle$, while, in some sense, it occupies
all those states for which $c_i \ne 0$. This is, perhaps, the most
intriguing feature of quantum mechanics; so to say, something can be
here and there simultaneously, without being here or being there
exactly.

There are many experiments in quantum mechanics demonstrating this
strange existence. Without going into technical details, let us
describe double-slit experiment in brief. It shows that some
micro system behaves like a wave during undisturbed propagation along
the interference device. Namely, by repeating the experiment many
times, as an accumulated result observer gets so-called interference
pattern. Its undulating character indicates that each system from the
ensemble went through both slits simultaneously, {\it i. e.}, that
systems were in state $\sum _{i=1}^2 c_i \vert a_i \rangle$ during
propagation. (Vector $\vert a_1 \rangle$ means `system went through the
first slit' and vector $\vert a_2 \rangle$ means `system went through
the second slit'.)

It is possible to construct slightly different experimental settings for
which each system passes through only one slit, despite of that both
are at disposal. System behaves like a particle then being localized in
a very small region around one of the slits. As was mentioned above, it
is impossible to predict through which of the slits system will go.
Measurement of path gives the answer to that question in particular
case. But, this measurement, on the other hand, destroys interference
pattern - it causes collapse of state: state (3) that describes
correctly ensemble before the measurement changes into state (4) after
the measurement.

Collapse or reduction of state is another novelty of quantum mechanics
closely related to superposition of states. It is discontinuous change
of state and occurs when quantum system interacts with classical one
(measuring device) and that in case when state of measured system is
superposition of the eigenstates of measured observable. Reduction is
transition from coherent to noncoherent mixture of involved states.
Potential existence of superposed properties before the collapse become
actual after it; before the collapse all systems of the ensemble are
in the same state, afterwards ensemble splits into subensembles of
systems in different states. Thorough analysis of the collapse one can
find in [3].

Superposition as a typical quantum phenomenon is hard to understand
because we are familiar with classical objects, not quantum, and our
minds operate according to the classical logic. If some particle like
system of our everyday experience can be in one of $\vert a_i \rangle$,
then that is all, it certainly cannot be in some superposition of these
states. With waves of classical world we do not have intuitive problems
since for these collective phenomena we have not noticed some sort of
spontaneous collapse. That is, our problems with superposition come
from the fact that reality of both $\vert a_i \rangle$'s and their
linear combinations have to be treated on equal footing. We are faced
with particle like objects, for which $\vert a_i \rangle$'s can be
observed and not their linear combinations and waves to which only
$\vert a_i \rangle$'s in linear combination could be attached, but
neither of them can appear separately.

How to understand superposition of states when the world we live
in is the world of macroscopic objects for which this concept does
not apply? How, then, to feel the novelty brought by quantum
mechanics? Is it possible to find some instructive example without
invoking physics of micro systems; example that will demonstrate
superposition and related topics in a way understandable even to
inexpert? These and similar questions can be summarized in the
following one: can fundamental elements of quantum mechanics be
found out of physics?

The answer could be, we believe, affirmative. Optimism is based on that
there is the whole world of symbols. It is true that, as macroscopic
material objects, they behave according to the laws of classical
mechanics, but their meaning need not to be constrained by this theory.
They could be organized in a way that corresponds to quantum logic.
This would be difficult, but it is not impossible. The art offers great
many possibilities for doing that since there are no limits on its
expressive power. The piece of art can lead us to some well organized
world which has nothing in common with the one of our everyday
experience.

What we shall try to show here is that this has already been done.
Precisely, we shall try to show that expression of Mona Lisa by
Leonardo da Vinci [4], Fig. 1, the famous smile, actually is
superposition of two different expressions. But, before going into
detailed elaboration, few remarks are in order.

First of all, and this is a commonplace, Leonardo was extremely gifted
person with extraordinary skills. Because of his aesthetic sensibility,
deep providence and patience in work, due to the investigations in
anatomy and awareness of laws of nature, especially optics, it is
impossible that Mona Lisa's expression and the whole painting, is
accidental or consequence of uncontrolled gesture. It is result of
precise intention and tremendous effort.

Second remark is that Leonardo was interested in combining things,
about which one can find beautiful lines in the novel Resurrected Gods
by Dimitry Sergeyevich Merezhkowsky. Juxtaposition of different or
contradictory entities, as it is well known, rises tension or
confrontation among them. So, the piece of art gets new quality; object
is no longer static and stable, it is in latent motion, has a sort of
vitality.

Finally, let us stress that we are not interested here in whom, if any,
was a person that posed for Leonardo. (Some believe that picture is
a kind of self-portrait since Gioconda has the artists scull.) We are
not interested in why Mona Lisa has such expression as well. (Doctor
Filippo Surano believes that Mona Lisa suffered from bruxism - an
unconscious habit of grinding the teeth; so the reason for smile is in
compulsive gnashing of teeth. On the other hand, Janusz Walek [5],
in order to prove that the smile is in accordance with the contemporary
manners, quotes Agnolo Firenzuola who advised women of Gioconda's time how
to smile in charming way.)  We only want to address the question why
Mona Lisa's smile is mysterious.

The whole collection of various impressions regarding Mona Lisa's smile
one can find in [6].  There is quotation of Th\'eophile Gautier's
opinion: `. . . but the expression, wise, deep, velvety, full of
promise, attracts you irresistibly and intoxicates you . . .' Ernst
Hans Gombrich [7] continues by saying that `What strikes us first
is the amazing degree to which Lisa looks alive . . . Like a living
being, she seems to change before our eyes and to look a little
different every time we come back to her. . . . Sometimes she seems to
mock at us, and then again we seem to catch something like sadness in
her smile. All this sounds rather mysterious and so it is, . . . '
This is in accordance with Giorgio Vasari's description, given in Lives
of the Artists: `The eyes have the brightness and moisture of the
living ones. . . . The opening of the mouth, . . . , seemed not to be
colored, but to be living flesh. . . . The picture is considered the
most splendid work, nearly alive.' Neuroscientist Margaret Livingstone
confirms this by claiming that the viewer sees Mona Lisa's face as
constantly changing, it has a `flickering quality - with smile present
and smile gone - which occurs as people move their eyes around Mona
Lisa's face.' Roy McMullen [6] explains this by claiming that ` . . .
the Mona Lisa is certainly a very undecided sort of creation . . . the
painting is self-contradictory not only in its details but also in its
message: the ambiguity is both a part of the subject and in a sense the
whole of it.'

Enigmatic, mysterious and incomprehensible are often used epithets
of Gioconda's remarkable portrait and her opaque, vague or simply
ineffable smile. Abandoning oneself, spectator goes through
perplexing transformation of impressions. Sensation that she is
smiling is instantaneously repelled by her melancholy;
cheerfulness becomes sadness and vice versa. Mona Lisa irritates
us for she is changing moods and this perpetual change is what
makes us to believe that she is alive. On the other hand, if
observer scrutinizes details, then striking dissimilarity among
elements indicating Gioconda's feelings appears. One can classify
them in distinguished sets: one pertaining to smiling, cheerful
Mona Lisa, one reflecting melancholy or disappointment and one
containing elements that connect previous producing complete and
realistic face.

For example, there is a shadow at the right corner of Gioconda's mouth,
oriented downwards and suggesting sadness, see Fig. 2. Beside it, there
is a dark area on the right cheek starting from the end of the lips and
going upwards at some angle. This, of course, designates cheerful mood.
So, these two shades certainly have opposite effects on spectator and
the picture is full of such narrative pieces. Instead of enumerating
them, let us point out that Leonardo has linked these contradictory
elements in a kind of dark and long strip. It consists of these shades
at the ends and a hardly noticeable spot that is approximately at the
same horizontal level as the mouth, being at its right. Purpose of this
spot is to integrate signs of unequal meaning in an indivisible entity,
making smooth transition among them without belonging completely to
any of these two.

One can proceed in this way by analyzing picture piece by piece: shadow
above her left eye designate sorrow, lightened region between temple and
left eye lifts up the cheek giving contribution to the opposite
impression {\it etc.}  We believe that much better than listing
cheerful and cheerless elements, is to present visually final result.
This is done by Fig. 3. On one side, all details spoiling the impression
that Gioconda is sad are covered, while, at the other side, excluded
are those that disturb us in seeing her cheerfulness. (Needless to say,
covering the original with white squares was the only intervention
taken.) Like damaged frescos, these faces, we believe, offer enough
material necessary for their understandings. First one shows Mona Lisa
in a state of calm contemplation on some unhappy events. This is
confirmed almost immediately, just after our imagination
straightforwardly interpolates residual pieces. The other one shows her
plain or maybe malicious smile; certainly, there is no dolor in her
eyes at this picture.

Before discussing these pictures from the point of view of physics, let
us comment original and try to explain how it was possible to
incorporate two distinguished expressions into single portrait. In
Leonardo's own words, given in one of his recommendations to painters
[8], `. . . laughing and weeping, . . .  are very similar in the motion
of mouth, the cheeks, the shutting of the eyebrows and the space
between them . . . ' So, for the great master, it was not {\it a
priori} impossible to combine even mutually exclusive emotions. But,
this similarity could only be necessary, definitively not sufficient
for designing such combination. Since Leonardo studied anatomy with
both artistic and scientific intention, he was capable to create
the most adequate face. (For instance, there are claims that Mona Lisa
had swellings on lower jaw, which strongly influenced distribution of
shades on her cheeks.) Then comes his treatment of light and
shadow. As stressed in [9], Leonardo avoided above all light which casts
a dark shadow. He preferred so-called diffuse light - the one that
comes from many sources, which are usually of low intensity. What is
more important, according to Kenneth Clark, for Leonardo shadow was an
adjunct of form. This means that delicate modifications of shape with
purpose of introducing and then balancing discrepancies could be
accomplished with a proper use of light and shadow. Closely connected
to this is the manner in which the painting has been executed. In [7]
it is said about this that `If the outlines are not quite so firmly
drawn, if the form is left a little vague, as though disappearing into
shadow, . . . impression of dryness and stiffness will be avoided. This
is Leonardo's famous invention, which Italians call sfumato - the
blurred outline and mellowed colors that allow one form to merge with
another and always leave something to our imagination.' And further `.
. . . what we call its expression rests mainly in two features: the
corners of the mouth and the corners of the eyes. Now it Are precisely
these parts which Leonardo has left deliberately indistinct, by letting
them merge into a soft shadow. That is why we are never quite certain
in which mood Mona Lisa is really looking at us.'

Regarding quantum mechanics and its terminology, system under
investigation is a female face. Measured observable is the mood of
portrayed person. Among many of its eigenstates, we are here interested
only in two of them, say sadness and cheerfulness. (Perhaps one can
find more adequate terms to name the appropriate moods of Mona Lisa,
but that is not important here for their purpose is just to refer to
the pictures given in Fig. 3.) How it looks like when the system is in
one or the other of these eigenstates we have tried to demonstrate by
Fig. 3. To be precise, there given pictures only indicate what are
Gioconda's pure states of sadness and cheerfulness. However, we find
them sufficiently suggestive - after few moments our imagination
provides complete image. (It would be hard task to finish them because
it has to be done in a manner that imitates Leonardo's, {\it i. e.}, it
demands skills which present author does not possess. Therefore, we
ignore the fact that complete and incomplete pictures of the same face
might be taken as different systems.)

When the ensemble of systems is in one of the eigenstates of the
measured observable, then one always obtains the same result after
measuring that observable. Exactly this happens with pictures of Fig.
3. Namely, if the first picture is under consideration, one has the
impression that Mona Lisa is sad while looking at the second one always
leads to the same conclusion - Mona Lisa is smiling. In each of these
two cases, ensemble consists of the picture we are looking at,
perceiving it dozens of times in a second. At one instant we notice
some detail on the picture and conclude that she is in particular mood,
then we look again, notice some other detail, make the same conclusion
and so on. Looking at the picture is nothing else but repetition of
measurement and the absence of doubt about the expressions at the end
means that there are no deviations, {\it i. e.}, all results of
observations coincide.

Since these two pictures have been extracted from the original, it is
plausible that the Gioconda's portrait is dichotomous. Her disquieting
smile is a mixture of two pure states and the question is whether it is
coherent or noncoherent one. We shall try to convince the reader that
it is coherent mixture - superposition. For that purpose, we firstly
have to propose some criterion or procedure how to distinguish coherent
from noncoherent mixture of the same states. (It is futile to try to
measure some other observable, which does not commute with the
considered one. All our sensual perceptions mutually commute;
otherwise, it would be ease to understand quantum mechanics.)

Difference between coherent and noncoherent mixtures is in that the
former are pure states, while the latter are not. For every pure state -
vector of the Hilbert space, there is one vector orthogonal to it if
the space is two-dimensional. This is not the case for mixed states in
such spaces. So, the procedure how to distinguish mentioned mixtures
could be the following: is it possible to find state that is orthogonal
to considered one or not? Here, this reads: is it possible to imagine,
or perhaps paint, portrait with expression opposite to the Mona Lisa's,
where opposite means as opposite as are those given in Fig. 3. We
believe that the answer is affirmative. It is possible, but its
realization demands   artist of talent and skills comparable to
Leonardo's.

That mysteriously crying Gioconda would be as provocative as is the
original since there would be some happiness in her dolorous eyes. And
this would be the only similarity; her undecided expression would not
be in any respect close to the Mona Lisa's, on the contrary.

However, since this way of convincing the reader that portrait shows
superposition of two moods rests on hardly feasible process, let us
propose the other one. As mentioned above, formal difference between
coherent and noncoherent mixture of the same states is in presence of
so-called off-diagonal terms in the former state. Due to them, this
state is idempotent, {\it i. e.}, pure. So, if Mona Lisa's expression
is coherent mixture, then there should be some off-diagonal elements.

What could be those off-diagonal elements? In order to answer this
question, let us start with Fig. 3. The pictures of this figure
undoubtedly display certain well-known moods. This means that these
pictures present Mona Lisa's pure states: first one shows Mona Lisa in
the pure state of sadness, say $\vert a_1 \rangle$, while the other
shows her in the pure state of cheerfulness, say $\vert a_2 \rangle$.
To have a portrait showing some pure state presumes that all its
narrative elements are in adequate position. Therefore, all details of
the first picture are in state $\vert a_1 \rangle$, while those of the
second are in $\vert a_2 \rangle$ or, in the operator form, $\vert a_1
\rangle \langle a_1 \vert$ and $\vert a_2 \rangle \langle a_2 \vert$,
respectively. So, diagonal elements are details of the portrait that
undoubtedly designate certain emotion or mood. Then, off-diagonal
elements $\vert a_1 \rangle \langle a_2 \vert$ and $\vert a_2 \rangle
\langle a_1 \vert$ can only be those details of the portrait that
somehow reflect two moods, those that are at one side related to the
first and at the other side to the second mood or those that connects
parts of opposite meaning. And Mona Lisa's portrait contain such
two-sided details, {\it e. g.}, mentioned dark spot beside the mouth.

All items of Leonardo's painting that are between signs of explicit
cheerfulness and explicit sadness are examples of these off-diagonal
elements. Connecting opposite, $\vert a_i \rangle \langle a_j \vert$
($i\ne j$) details make Gioconda's expression smooth, they provide
continuous flow from parts of one to those of the other mood. Due to
them, Mona Lisa is in pure state. We take her portrait as realistic
just because of these off-diagonal elements; they make her expression
somehow compact, persistent or impenetrable.

Mona Lisa is in pure state that is unusual combination of cheerfulness
and sadness, {\it i. e.}, her state is coherent mixture or
superposition of cheerfulness and sadness - it is (3) and not (4).
Without $\vert a_i \rangle \langle a_j \vert$ ($i\ne j$) elements, Mona
Lisa's portrait would be collection of unrelated dissonant pieces with
noticeable discontinuity among them. It would not be irresistible at
all; no one would be confused or irritated by such senseless
assemblage. Literally, it would be noncoherent mixture. (It is easy to
make such picture by gluing together pieces of two photos of the same
person in different moods.)

Finally, if one has agreed that Gioconda's portrait is superposition of
two states, then one might wonder what happens when we look at her. If
we do it unintentionally, at one instant we notice some suggestive
detail and make conclusion about her mood. That is, we find result of
measurement. At next moment, we repeat observation and make new
conclusion and so on. We notice different parts of the picture
randomly, so our impression varies; it is unpredictable what shall draw
our attention at particular moment. In this way her expression is being
projected either to cheerful or cheerless state, depending on what we
have just seen. While we look at her, Mona Lisa's expression
spontaneously collapses from (3) into one of two possible states - to
watch her means to make measurement of her mood and this causes
reduction of the state she is in. States like (4) in
adequate way summarize our opinions about which we may say that more
often we find Gioconda clearly smiling than being sad, so $w_1 < w_2$.
With state such is (4) we can represent ensemble of our impressions
and what is astonishing is that, in the aggregate, we end with Mona
Lisa in mixed state (noncoherent mixture) despite of the fact that we
are confronted with static object.

Our attempt to avoid these reductions results in puzzle. If we try
to catch Gioconda's real mood without rushing into conclusion, we
notice $\vert a_1 \rangle \langle a_1 \vert$, then $\vert a_1
\rangle \langle a_2 \vert$ leads us to $\vert a_2 \rangle \langle
a_2 \vert$ and $\vert a_2 \rangle \langle a_1 \vert$ takes us
back. Instead to conquer her, we do not know what to think about
her at all; by trying to comprehend her expression {\it in toto},
we actually discover state (3). And we are less satisfied with it
than with changeable impression since we are unable even to
entitle this state. What we see is not just beyond our experience,
this expression is graspable in its true meaning, it evades our
final judgment leaving us in confusion. And what we have to do is
not to look for trivial resolution of the problem, with experience
of Mona Lisa's smile we should transcend limits of our cognitive
power.

To conclude, enigmatic smile of Mona Lisa shows that, beside the
classical logic, the underlying logical structure of quantum mechanics
can be implemented in creating the piece of art as well. It is an
example of superposition of two different states. Leonardo's
masterpiece is a visual illustration of coherent mixture (3), it could
be an emblem of these mixtures. On the other hand, the only way to
understand entirely Mona Lisa's expression and our consecutive
impressions is by referring to quantum mechanics. If we do not employ
subtle formalism of this theory, it seems to be impossible to find
convincing and full explanation of our mental journey. Since our minds
operate according to classical logic, just described processes are
normal reactions; we have to simplify her expression by projecting it
to well known ones or we shall be driven along the M\"obius strip.

\end{document}